# Diabetic Retinopathy Diagnosis based on Convolutional Neural Network

**Mohammed hamzah abed[1], Lamia Abed Noor Muhammed[1], Sarah Hussein Toman[2]**
[1]Faculty of Computer Science and Information Technology, University of Al-Qadisiyah, Iraq
[2]Roads and Transport Department, College of Engineering, University of Al −Qadisiyah, Iraq

**ABSTRACT**

Diabetic Retinopathy DR is a popular disease for many people as a result of age or the diabetic, as a result, it can cause blindness. therefore, diagnosis of this disease especially in the early time can prevent its effect for a lot of patients.  To achieve this diagnosis, eye retina must be examined continuously. Therefore, computer-aided tools can be used in the field based on computer vision techniques. Different works have been performed using various machine learning techniques. Convolutional Neural Network is one of the promise methods, so it was for Diabetic Retinopathy detection in this paper. Also, the proposed work contains visual enhancement in the pre-processing phase, then the CNN model is trained to be able for recognition and classification phase, to diagnosis the healthy and unhealthy retina image. Three public dataset DiaretDB0, DiaretDB1 and DrimDB were used in practical testing.  The implementation of this work based on Matlab- R2019a, deep learning toolbox and deep network designer to design the architecture of the convolutional neural network and train it. The results were evaluated to different metrics; accuracy is one of them. The best accuracy that was achieved: for DiaretDB0 is 100%, DiaretDB1 is 99.495%  and DrimDB is 97.55%.

*Corresponding Author:*
Mohammed Hamzah Abed,
Department Computer Science,
Al-Qadisiyah University,
Al-Dewaniyah, Iraq.
Email: mohammed.abed@qu.edu.iq

## 1. INTRODUCTION

One of the most common eye's diseases that is Diabetic Retinopathy DR. It is one of the main causes the blindness in a working-age population [1], afflict the people who are suffering from diabetes for many years. According to the World Health Organization WHO, around 1.6 million died due to diabetes in 2016 [2]. While 425 million people are living with diabetic all over the world, the big number of them are suffering from diabetic retinopathy [1]. In case of the most developing country like USA 40% of 29.1 million diabetic patients suffering from DR in different stages this ratio in 2018 [1], as well as there is a study from another country that is Iraq, they were collecting information from three clinics centers according to the period from 1 September to 30 November 2007 the ratio of DR was 76.1 from 18612 patients [3] and the number of cases still have been increased, another study in 2015 of type 2 diabetic 33.1% of patient had Retinopathy[4]. Diabetic retinopathy is divided into two stages: non-proliferative (NPDR) and proliferative diabetic retinopathy (PDR). the early stage of DR is NPDR. Through early stage, increased vascular permeability and capillary occlusion are observed in the retinal vasculature [5]. Retinal pathologies including micro aneurysms, hemorrhages and hard exudates can be detected by fundus photography so they can be used to diagnosis DR at an early stage. The detection of the disease at an early stage is more important in order to treat the patient, so this need for examining the retina in continuously, however with the advances in the vision computer technology, it can be used as a tool for detection automatically. This tool has its characteristics; it can be available in any time to perform the examination of an eye without need for the physician so it can give the early detection for this disease and in result, the patient takes the treatment in proper time. Many approaches have been suggested to diagnosis DR disease, Automatic DR diagnosis system based on deep learning technologies have achieved a promising accuracy than other techniques of computer vision [6][7]. In this research work will focus on deep learning technique to detect and diagnose the DR of retina images. The remainder of this paper organizes as follows. Section 2 presents the literature survey and the latest techniques used to detect and diagnose. Section 3 presents the methodology and the proposed model. Section 4 discuss the experimental result and analysis and dataset that are used to test the performance of accuracy. Finally, conclusion and future work are presented.

## 2. LITERATURE SURVEY

Diabetic Retinopathy detection automatically is important for early treatment, however, this disease causes the blindness around a world, based on the result of WHO [2]. Therefore, this is a promising field for many of the researchers, have worked in suggestions and developing. A summary of the literature, interested in Diabetic Retinopathy analysis and diagnosis would be presented, according to Morphological that has used in disease detection and the methods have been adopted [8][9] [10]. Nidhal K ans Enas suggested automatic early diagnosis of diabetic retinopathy, based on an abnormal indication from exudates, bleeding spots, blood vessels diameter and blood vessels tortuosity; using simple comparison algorithm with fundus images that collected from 100 patients with diabetic 10 of them suffering from eyes issues causes by diabetic, the accuracy was 100%[11]. Sopharak et al. designed an automatic model for diabetic detection based on the morphological concept through intensity analysis of retina to detect exudates [10]. in 2012 the researchers Giancardo et al used green channel of RGB color space and the intensity of the channel from eye's retinal images to detect the exudate [12]. In another hand Sánchezet al. suggested technique based on a statistical model and clustering for dynamic thresholding of the exudate pixels [13]. B_alint BORSOS et al. presented an automatic system for detection soft and hard exudate in order to diagnosis DR, using different steps such as segmentation and normalization then applied ANN to identify the pixel that belongs to exudate [14]. Deep learning techniques have achieved promising results in computer vision problems [15], especially medical image analysis [16] [17]. P. Prentǎsiće and S. Lončarić worked on design a convolutional neural network (CNN) with ten layers and trained it to detect the exudate of the RGB color retina images [18]. Parham Khojasteh et al suggest a novel color space model of fundus images for automatic exudate detection system [19]. The researcher used CNN to assess the accuracy of performance of different color space models. Khojastehet al. performed the comparison of the accuracy performance of different methods of Deep Learning for exudates detection and they obtained an accuracy of 89.1% for the CNN model [20]. the researcher in [21] used model of automatic optic disk to determine eye's health based on entropy factor. The researcher in [22] proposed a hybrid statistical framework for DR detection was suggested that combines the probabilistic SVM-based kernel with scaled Dirichlet distributions. all literature indicates that deep learning and convolutional neural network CNN are promising for automatic diabetic diagnosis based on color retina images.

## 3. MATERIALS AND METHODS

In this section, it is explained the proposed model system and materials of research , as well as the architecture of convolutional neural network.

### 3.1. Diabetic Retinopathy Detection

The retinal function requires the presence of blood-retinal tissues, where the tissues of blood are spread. The abnormal condition of the sugar level in the retinal blood vessels leads to increased vascular permeability and lipoprotein plaque formation occur early in diabetic retinopathy [23], The buildup of lipoprotein exudates leads to macula edema as shown in Figure 1, negatively affecting vision so it is an early sign of diabetic retinopathy, which can be considered as a basic element of diabetic retinopathy. Later, the abnormal retinal blood vessels may breakdown into the form of microvascular networks, which is called retinal neovascularization [24]. In addition to these signs, abnormalities including cotton wool spots, hemorrhages, exudates which lead to non-reversible blindness and vision impairment.

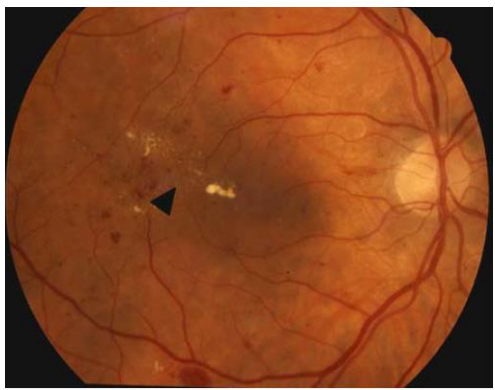

Figure 1 The black arrow points out a lipoprotein plaque located in the macula

To help in detection diabetic, fundus photographs is the back of the eye retinopathy may be taken because these photos make it easier for the physician to monitor DR signs, examining the retinal images are exploited in analyzing various aspects and stages of retinopathy. The computer vision tools have been used in the detection of Diabetes Retinopathy automatically through detection various features and stages of and can be referred to the specialist accordingly for intervention [25]. The availability, speed and high performance of these tools making the effective screening of Diabetic Retinopathy patients.

### 3.2. Convolutional Neural Networks

Deep learning, is an approach of AI. Specifically, it is a type of machine learning, a technique that allows computer systems to improve with experience and data. It achieves great power and flexibility by learning to represent the world as a nested hierarchy of concepts. It has been successfully used in commercial applications since the 1990s. Deep learning algorithms have completely changed our perception of information processing through different applications such as; NLP, computer vision, speech and audio, social network analysis, and healthcare, this includes application area the computer-aided diagnosis [26]. Some key enabler deep learning algorithms such as generative adversarial networks, convolutional neural networks, and model transfers [27]. Convolutional Neural Network (CNN model) is influenced by the human visual system [28]. CNN architecture consists of multi-layer. Low-level features and attributes can have extracted from earlier layer for example edge detection. And the higher layer or final fully connected layer gives the object features of the entire image [15] [29] [30]. Convolutional Layer is central of CNN [31] [32]. It takes a filter(kernel) and pass it through all the points in the image (Input) and passing at any point into a single position (Output). Pooling layer down-sampling of an image, It takes sub-samples of convolutional layer output and produces a single output. Then it uses different pooling techniques such as max pooling, mean pooling, average pooling etc. Later, fully connected layers take input from all neurons in the previous layer and perform operation with an individual neuron in the current layer to generate output.

### 4. Practical Work and Results

The proposed model of automatic Diabetic Retinopathy's detection, are dividing into three phases. Starting from preparing the data, then extract the features of the entire images based on CNN, finally classification phase to recognize healthy and unhealthy retina image. Figure 2, shows the block diagram of the proposed model schema.

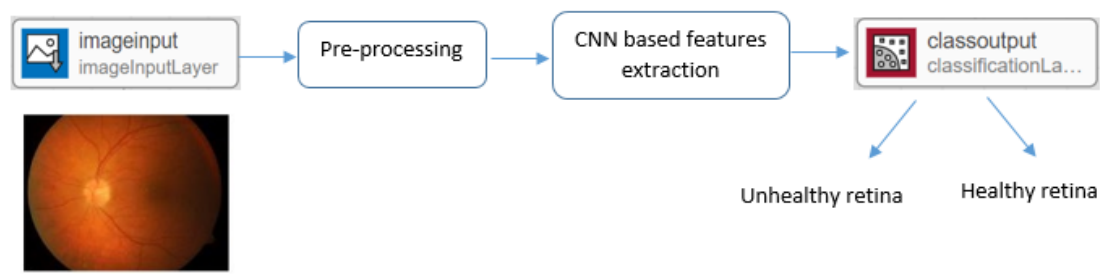

Figure 2 proposed model of DR

### 4.1 Preprocessing

The first phase of the proposed model is preparing the data, by uniform the size of three different datasets to be suitable for CNN architecture. The retina images first are extracted from the background by detecting the boundary of the retina. Enhance the images to show blood vessels and make it more visible using Contrast-limited adaptive histogram equalization (CLAHE). Figure 4 shows the original image and the enhanced retina image.

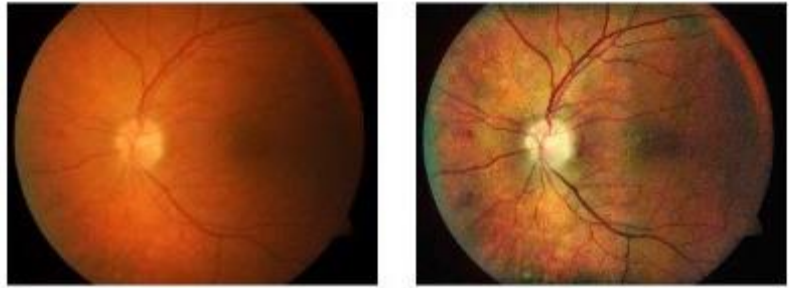

(a) (b)

Figure 4 (a) original retina image (b) enhanced image

### 4.2. Deep Network Design (CNN architecture)

the proposed CNN architecture consists of three phases of a sequence of convolution layer, normalization layer with epsilon 0.00001, then activation function Rectified Linear Unit (ReLU) and max-pooling layer. The features maps are calculated within convolution layer are normalized then based on ReLU for transforming the sum of weighted input into the activation of the node of the output of the current phase, this kind of activation function are easy to train and achieve a good performance. In the pooling layer, the maximum or average of the neighboring values to the feature maps are calculated to speed up the processes and reduce the size. These phases are repeating three times in our model, then first fully connected layer are learned based on drop out with probability 0.3, and the second fully connected layer are learned by soft-max function. Figure 5 shows the CNN architecture of the proposed model.

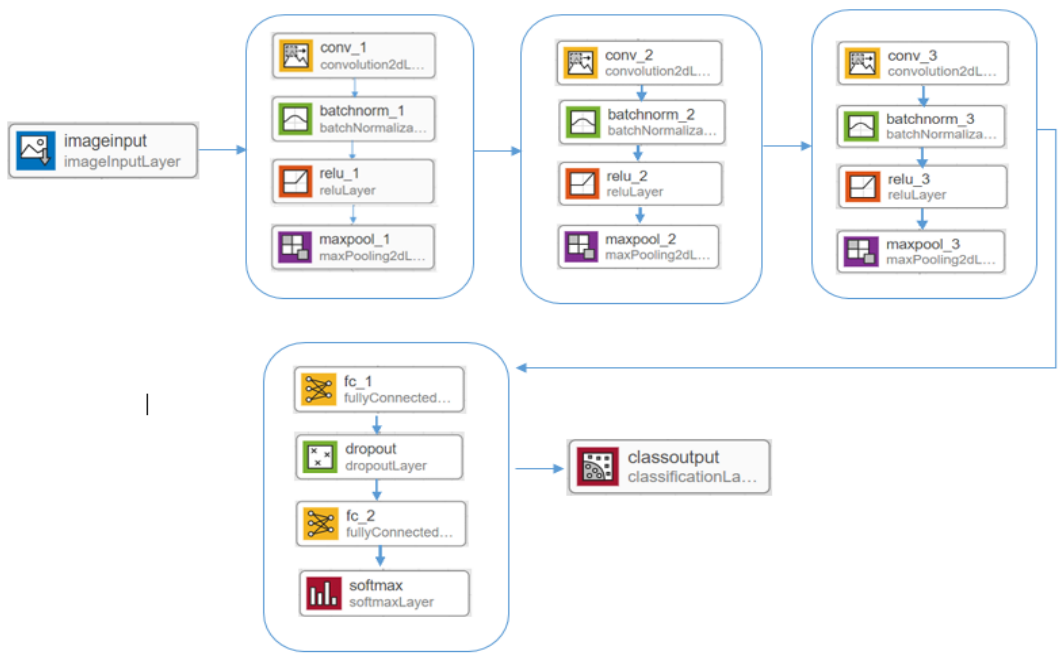

Figure 5 CNN architecture of proposed model

Table 1 shows the configuration of suggested CNN architecture, that suggested in proposed model.

Table 1 CNN architecture configuration

| Layer | Configuration | Description |
| --- | --- | --- |
| Conv_1 | Filter size 3,3<br>Number of filter 8<br>Stride 1,1 | Convolutional layer 1 |
| batchnorm_1 | Epsilon 0.00001 | Batch Normalization layer 1 |
| relu_1 | | Activation function 1 |
| maxpool_1 | Pool size 2,2<br>Stride 2,2<br>Padding 0,0,0,0 | Pooling layer 1 |
| conv_2 | Filter size 3,3<br>Number of filter 16<br>Stride 1,1 | Convolutional layer 2 |
| batchnorm_2 | Epsilon 0.00001 | Batch Normalization layer 2 |
| relu_2 | | Activation function 2 |
| maxpool_2 | Pool size 2,2<br>Stride 2,2<br>Padding 0,0,0,0 | Pooling layer 2 |
| fc_1 | Output size 2<br>Weight learn rate factor 1 | Full connected layer 1 |
| dropout | Probability 0.3 | Normalization |
| fc_2 | Output size 2<br>Weight learn rate factor 1 | Full connected layer 2 |
| Softmax | | |
| classoutput | Two output | Output |

### 4.3. Experimental result

The proposed model is tested on three different datasets, 80% of the retina images are selected for training and the 20% is for testing, a random selection of training images set for better evaluation. The experiments were performed on a computer with Matlab 2019a in Windows 10 64 bit operating system environment. The training options of the proposed CNN were using sgdm and learning rate schedule piecewise, with drop factor 0.2 to reduce the learning rate by a factor every 5 epochs, Set the maximum number of epochs for training to 20, shuffle each epoch

1. Images Dataset

    The data-sets that used in the proposed model are: DIARETDB0 [33], DIARETDB1 [34][35] and DRIMDB [36].

    DIARETDB0 is a first dataset were used to test the proposed model. it is consisting of 130 color fundus images of which 20 are normal and 110 contain signs of the diabetic retinopathy. DIARETDB0 and DIARETDB1 databases were taken using a digital fundus camera with a viewing angle of 50 °and a resolution of 1500 ×1152 at 24 bits. The DIARETDB1 consisting of 89 color fundus images, 84 images of them contain sign and the 5 considering as a normal which do not contain any sign. The third dataset is DRIMDB consisting of three classes good, bad and outlier the total number of images is 216.

2. Classification result and Analysis

    the proposed model based on CNN is tested on three different datasets, the datasets randomly divided into 80% for training and the reminder 20% for testing. During the training option, the validation is testing for each epoch, table 2 shows the validation accuracy during a training phase.

Table 2 validation accuracy based on each epoch

| Dataset | Epoch | Validation Accuracy |
|---|---|---|
| DIARETDB0 | 1 | 62.67% |
| | 13 | 100% |
| | 20 | 100% |
| | 1 | 57.9% |
| DIARETDB1 | 13 | 99.89% |
| | 20 | 99.1% |
| DRIMDB | 1 | 68% |
| | 13 | 95.1% |
| | 20 | 100% |

The training accuracy and loss function of selected dataset DIARETDB1 as shown in figure 6, during training phase.

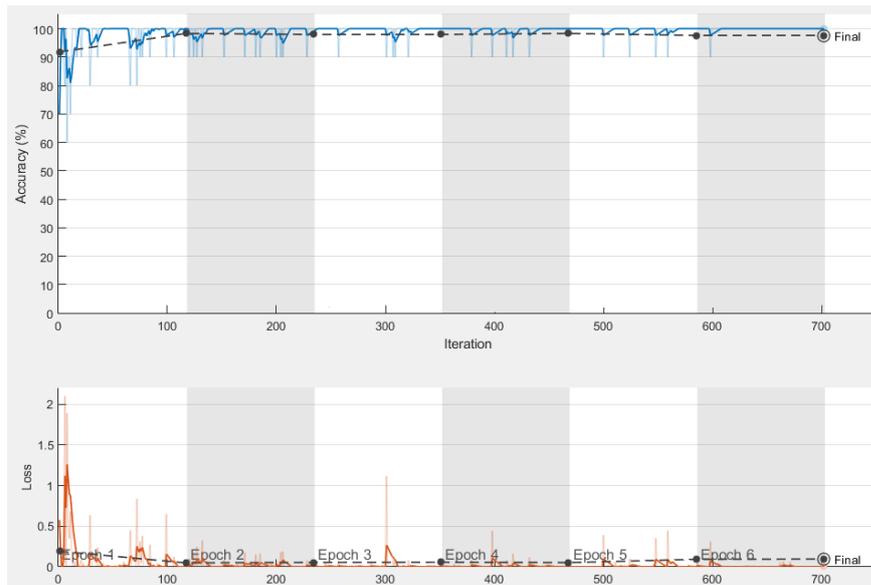

Figure 6 accuracy and loss function of DIARETDB1

In comparing the results obtained in this paperwork with some related work for the same datasets is shown in table 3. It is obvious, good results are achieved so we can these results are promising to evolve the proposed model to be applied with other datasets.

Table 3 summary of accuracy comparing with different methods

| Paper proposed work | Dataset | Accuracy |
|---|---|---|
| Zeljkovi ´c, V and elt.[37] | DIARETDB0 - - | 93.16 |
| Kemal Adem 2018 [28] | DIARETDB0 DIARETDB1 DRIMDB | 100 % 99.2% 100% |
| Proposed model based on CNN 2020 | DIARETDB0 DIARETDB1 DRIMDB | 100 99.495 97.55 |

The proposed model trained separately for each dataset and test it on the different test part that belongs to each dataset. The success classification ratio as shown above in Table 3. The proposed model achieved good result than other in [37], and achieved equal success accuracy in DIARETDB0 , as well as higher result in DIARETDB1 ,and less success accuracy in the third dataset DRIMDB.

## 5. CONCLUSION AND FUTURE WORK

One of the most common eye's diseases is Diabetic Retinopathy. In this study proposed convolutional Neural Network is suggested for classification healthy and unhealthy retina images based on the infection of blood vessels. The suggested model consists of three phases pre-processing, features extraction based on CNN and classification. The CNN architecture model contains 17 layers starting from the input layer to the classification layer. In the future, the model will be improved and trained on more dataset to be able to recognize more disease to help the physician in the clinic and hospitals.

**ACKNOWLEDGEMENTS**

We would like to thank Al-Qadisiyah University for support.

**BIOGRAPHIES OF AUTHORS**

| | |
|---|---|
| 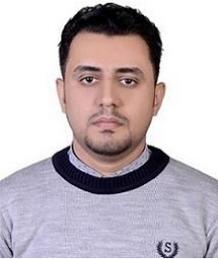 | Mohammed Hamzah Abed , recevied the B.Sc degree in Computer Science from University of Al-Qadisiyah , Iraq 2008, M.Sc degree in Computer Science from B.A.M. University ,india 2011.currently, I work as a lecturer at the Department of Computer Science, University of Al-Qadisiyah.I'm doing research in Medical Image Processing and machine learning. my current work is Medical image analysis based on deep learning |
| **Lamia Abed Noor Muhammed** | Assisstant Professor in Computer Science, she is working as an academic staff at Al-Qadisiyah University, and focusing on machine learning and data mining.<br>B.Sc ,M.Sc,PhD in Computer Science. |
| **Sarah Hussein Toman** | Lecturer in Computer Science , she is working as an academic staff at Al-Qadisiyah University,College of Enginerring , my major research focusing on cloud computing and web mining and web technology<br>B.Sc ,M.Sc, in Computer Science. |